# On the origin of the metallic and anisotropic magnetic properties of $Na_xCoO_2$ ($x \approx 0.75$)

M.-H. Whangbo* and D. Dai

Department of Chemistry, North Carolina State University, Raleigh, North Carolina 27695-8204

R. K. Kremer*

Max-Planck-Institut für Festkörperforschung, Heisenbergstrasse 1, D-70569 Stuttgart, Germany

_________________________________________________________________

E-mail: mike_whangbo@ncsu.edu (M.-H. W.), rekre@fkf.mpg.de (R. K. K)

**Abstract**

Non-stoichiometric $Na_xCoO_2$ (0.5 < x < 1) consists of $CoO_2$ layers made up of edge-sharing $CoO_6$ octahedra, and exhibits strongly anisotropic magnetic susceptibilities as well as metallic properties. A modified Curie-Weiss law was proposed for systems containing anisotropic magnetic ions to analyze the magnetic susceptibilities of $Na_xCoO_2$ (x ≈ 0.75), and implications of this analysis were explored. Our study shows that the low-spin $Co^{4+}$ ($S = 1/2$) ions of $Na_xCoO_2$ generated by the Na vacancies cause the anisotropic magnetic properties of $Na_xCoO_2$, and suggests that the six nearest-neighbor $Co^{3+}$ ions of each $Co^{4+}$ ion adopt the intermediate-spin electron configuration thereby behaving magnetically like low-spin $Co^{4+}$ ions. The Weiss temperature of $Na_xCoO_2$ is more negative along the direction of the lower g-factor (i.e., $\theta_{//} < \theta_{\perp} < 0$, and $g_{//} < g_{\perp}$). The occurrence of intermediate-spin $Co^{3+}$ ions surrounding each $Co^{4+}$ ion account for the apparently puzzling magnetic properties of $Na_xCoO_2$ (x ≈ 0.75), i.e., the large negative Weiss temperature, the three-dimensional antiferromagnetic ordering below ~22 K, and the metallic properties. The picture of the magnetic structure derived from neutron scattering studies below ~22 K are in apparent conflict with that deduced from magnetic susceptibility measurements between ~50 − 300 K. These conflicting pictures are resolved by noting that the spin exchange between $Co^{3+}$ ions is more strongly antiferromagnetic than that between $Co^{4+}$ and $Co^{3+}$ ions.

**1. Introduction**

The crystal structure of $NaCoO_2$ consists of $CoO_2$ layers made up of edge-sharing $CoO_6$ octahedra with Na atoms intercalated in between the $CoO_2$ layers.[1-3] Stoichiometric $NaCoO_2$ is semiconducting, while non-stoichiometric $Na_xCoO_2$ ($0 < x < 1$) is metallic.[4] Since the discovery of superconductivity below ~4.5 K in $Na_{0.3}CoO_2 \cdot 1.4H_2O$,[5] the unhydrated compound $Na_xCoO_2$ ($0 < x < 1$) has received much attention.[6-11] To understand charge and spin ordering in the sodium-deficient compound $Na_xCoO_2$, it is crucial to know how the Na atoms are arranged in $Na_xCoO_2$.[8,11] The magnetic properties of $Na_xCoO_2$ ($0.5 < x < \sim 0.75$) show a Curie-Weiss behavior with a large negative Weiss temperature.[6,9,10] In particular, the magnetic susceptibilities measured for single-crystal samples of $Na_xCoO_2$ show a strong anisotropy;[7,9,10] the magnetic susceptibility $\chi_{//}$ parallel to the c-direction (i.e., the //c-direction, which is perpendicular to the $CoO_2$ plane) is smaller than the magnetic susceptibility $\chi_\perp$ perpendicular to the c-direction (i.e., the $\perp$c-direction, which is parallel to the $CoO_2$ plane),[9,10] and the $\chi_\perp$ versus $\chi_{//}$ curves show a very good linear relationship in the temperature region of $50 - 250$ K.[9]

Chou *et al.*[9] and Sales *et al.*[10] analyzed the parallel and perpendicular magnetic susceptibilities $\chi_i$ ($i = //, \perp$) of $Na_{0.75}CoO_2$ using the modified Curie-Weiss law

$$\chi_i = \chi_i^0 + \frac{C_i}{T - \theta_i}, \qquad (1)$$

where $\chi_i^0$ represents temperature-independent contributions, which include the Van Vleck paramagnetism and the core diamagnetism. The fitting parameters, $\chi_i^0$, $C_i$ and $\theta_i$, resulting from these analyses are somewhat puzzling. The Curie constant of a magnetic system is related to its effective moment at temperatures sufficiently high that the spin arrangement is random. Thus one might expect $C_{//} \approx C_{\perp}$, but $C_{\perp}$ is much larger than $C_{//}$ (e.g., $C_{\perp}/C_{//} \approx 2.44$ for $Na_{0.75}CoO_2$).[9] In general, a lower magnetic susceptibility is expected for the direction with a more negative Weiss temperature, but $\theta_{\perp} < \theta_{//} < 0$ despite that $\chi_{\perp} > \chi_{//}$ at a given temperature.[9,10] This situation arises because $C_{\perp} >> C_{//}$. In general, the temperature-independent contribution $\chi_i^0$ should be small compared with $\chi_i$, but the fitted $\chi_i^0$ is nearly comparable to $\chi_i$ in magnitude (e.g., $\chi_{//}^0/\chi_{//} \approx 0.41$ at 50 K, and 0.61 at 250 K, in $Na_{0.75}CoO_2$).[9] In addition, $\chi_i^0$ is strongly anisotropic (e.g., $\chi_{\perp}^0/\chi_{//}^0 = 0.57$ for $Na_{0.75}CoO_2$).[9] The core diamagnetism cannot be anisotropic, and the Van Vleck temperature-independent paramagnetism is often of the same order of magnitude as the diamagnetism, but of opposite sign. Thus, there is no compelling reason why $\chi_i^0$ should be so large and strongly anisotropic.

The nature of the magnetic structure of $Na_xCoO_2$ ($x \approx 0.75$) emerging from the neutron scattering [11-13] studies is in apparent conflict with that deduced from the magnetic susceptibility studies.[9,10] The neutron scattering studies of $Na_{0.82}CoO_2$ [11] and $Na_{0.75}CoO_2$ [12] carried out below 2 K led to the conclusion that $Na_xCoO_2$ ($x \approx 0.75$) has a type-A antiferromagnetic (AFM) structure, i.e., the spin exchange within each $CoO_2$ layer is ferromagnetic (FM) while that between adjacent $CoO_2$ layers is strongly AFM. Furthermore, FM spin fluctuations within the $CoO_2$ layers exist up to 200 K.[13] This finding is quite surprising because the Weiss temperatures $\theta_{\perp}$ and $\theta_{//}$ are both strongly negative between 50 – 250 K, which indicates strong predominant AFM spin exchange. If the spin exchange between adjacent $CoO_2$ layers were to occur through the Co-O…O-Co super-superexchange paths with very long interlayer O…O distances, strong AFM interlayer exchange interactions as the source of a negative Weiss temperature can be ruled out.[14]

In the present work we address the apparently puzzling observations summarized above. In what follows we propose a modified Curie-Weiss law, which allows one to describe magnetic systems composed of anisotropic magnetic ions. We then report results of our analysis of the anisotropic magnetic susceptibilities of $Na_{0.78}CoO_2$ measured for a single crystal sample on the basis of the modified Curie-Weiss law. For the sake of comparison, we also analyze the anisotropic magnetic susceptibilities of $Na_{0.75}CoO_2$ reported by Chou *et al.*[9] and Sales *et al.*[10] Then we probe implications of our results concerning the apparently puzzling magnetic properties of $Na_xCoO_2$ ($x \approx 0.75$).

## 2. Anisotropic character of a low-spin $Co^{4+}$ ($d^5$) ion at an octahedral site

The $Co^{3+}$ ($d^6$) ions of $NaCoO_2$ have the low-spin (LS) configuration $(t_{2g})^6$ and hence are diamagnetic and semiconducting.[4] In a non-stoichiometric compound $Na_xCoO_2$ ($0.5 < x < 1$), the charge neutrality requires the presence of $Co^{4+}$ ($d^5$) ions in mole fraction $(1-x)$. To explain the metallic property of a non-stoichiometric $Na_xCoO_2$,[4] one might suppose the delocalization of the charges of the $Co^{4+}$ ions leading to a partially empty $t_{2g}$-block bands. However, this picture does not explain the magnetic properties of $Na_xCoO_2$. To understand the anisotropic magnetic properties of $Na_xCoO_2$, it is necessary that the charge of a $Co^{4+}$ ion remains localized (see Section 5 for further discussion of the origin of the metallic properties of $Na_xCoO_2$). In principle, a $Co^{4+}$ ($d^5$) ion at an octahedral site may adopt the high-spin (HS) configuration $(t_{2g})^3(e_g)^2$ ($S = 5/2$), the intermediate-spin (IS) configuration $(t_{2g})^4(e_g)^1$ ($S = 3/2$), or the LS configuration $(t_{2g})^5$ ($S = 1/2$). A number of studies suggest that the $Co^{4+}$ ($d^5$) ions of $Na_xCoO_2$ adopt the LS configuration.[15] The latter has an important consequence. The extensively studied layered magnetic compound $CoCl_2$, which is made up of edge-sharing $CoCl_6$ octahedra containing HS $Co^{2+}$ ($d^7$) ions, exhibits anisotropic magnetic susceptibilities.[16] The electron configuration $(t_{2g})^5$ of a LS $Co^{4+}$ ion in $Na_xCoO_2$ is the same as the electron configuration $(t_{2g})^5(e_g)^2$ of a HS $Co^{2+}$ ion in $CoCl_2$, as far as the electron occupation of the $t_{2g}$ levels is concerned. The magnetic anisotropy of $CoCl_2$ arises primarily from the $(t_{2g})^5$ configuration and is related to how the ground electronic state associated with this configuration is modified by the crystal field, spin-orbit interaction and Zeeman effects.[16,17] As in the case of the $CoCl_6$ octahedra of $CoCl_2$, the $CoO_6$ octahedra of $NaCoO_2$ are slightly axially-flattened ($\angle$O-Co-O = 96.2°).[3] Therefore, the LS $Co^{4+}$ ions of $Na_xCoO_2$ ($0.5 < x < 1$) should exhibit anisotropic magnetic properties just as do the HS $Co^{2+}$ ions of $CoCl_2$, for which the g-factors along the //c- and $\perp$c-directions were estimated to be $g_{//}$ = 3.38 and $g_\perp$ = 4.84.[16] To a first approximation, these g-factors should be similar to those of LS $Co^{4+}$ ions in $Na_xCoO_2$.

## 3. Curie-Weiss law for anisotropic magnetic systems

For our discussion it is necessary to briefly review the mean-field treatment of Weiss for a magnetic system composed of isotropic magnetic ions.[18] A given spin $\hat{S}$ interacting with an external magnetic field $\vec{H}$ the Zeeman energy is described by the Hamiltonian $\hat{H} = g\beta\hat{S}\cdot\vec{H}$, where $g$ is the electron g-factor and $\beta$ is the Bohr magneton. In an extended magnetic system, a given spin also interacts with the magnetic field generated by its neighbor spins. Assuming predominant interaction with nearest neighbor (NN) spins, the effective Hamiltonian in mean field approximation is written as

$$\hat{H} = g\beta \hat{S} \cdot \vec{H} + (zJ\vec{M} / Ng\beta) \cdot \hat{S}, \tag{2}$$

where $z$ is the number of NN spin sites interacting with the given spin through the spin exchange parameter $J$, $\vec{M}$ is the molar magnetization, and $N$ is Avogadro's number. Eq. 2 can be rewritten in terms of the effective magnetic field $\vec{H}^{\text{eff}}$ acting on a given spin as

$$\vec{H}^{\text{eff}} = \vec{H} + zJ\vec{M} / Ng^2\beta^2. \tag{3}$$

When $\vec{H}^{\text{eff}}$ is small, $\vec{M} = \chi\vec{H}^{\text{eff}}$, where $\chi$ is the molar magnetic susceptibility. This relation and Eq. 3 lead to

$$\vec{M} = \frac{\chi\vec{H}}{1 - (zJ/Ng^2\beta^2)\chi}. \tag{4}$$

The molar magnetic susceptibility that is actually measured, i.e., $\chi_i^{\text{eff}} = \vec{M} / \vec{H}$, is then expressed as

$$\chi_i^{\text{eff}} = \frac{\chi}{1 - (zJ/Ng^2\beta^2)\chi}. \tag{5}$$

By replacing $\chi$ with $C/T$ (the Curie law),

$$\chi_i^{\text{eff}} = \frac{C}{T - (zJ/Ng^2\beta^2)C} = \frac{C}{T - \theta}. \tag{6}$$

Given the Curie constant $C = Ng^2\beta^2 S(S+1)/3k_{\text{B}}$, the Weiss temperature $\theta = zJ\,S(S+1)/3k_{\text{B}}$ does not depend on the g-factor.

For a system composed of anisotropic magnetic ions, it is necessary to modify the above treatment. Labeling the g-factors of an anisotropic magnetic ion of a layered compound along the //c- and ⊥c-directions as $g_{//}$ and $g_{\perp}$, respectively, the magnetic susceptibility $\chi_i^{\text{eff}}$ measured with the magnetic field $H_i$ ($i = //, \perp$) can be written as

$$\chi_i^{\text{eff}} = \frac{C_i}{T - (zJ/Ng_i^2\beta^2)C_i} \tag{7a}$$

$$= \frac{C_i}{T - \theta_i}, \qquad (i = //, \perp) \tag{7b}$$

from which we find

$$1/\chi_i^{\text{eff}} = (1/C_i)\,T - \theta_i/C_i \qquad (i = //, \perp) \tag{8}$$

As will be shown in the next section for $Na_xCoO_2$ (x $\approx$ 0.75), in the temperature region (i.e., $50 - 300$ K) where the $1/\chi_i^{\text{eff}}$ vs. $T$ plots are linear, the slope of the $1/\chi_{//}^{\text{eff}}$ vs. $T$ plot is nearly the same as that of the $1/\chi_{\perp}^{\text{eff}}$ vs. $T$ plot for $Na_xCoO_2$ (x $\approx$ 0.75). This implies that the Curie constants $C_{//}$ and $C_{\perp}$ are approximately equal. Thus, one may write

$$C_{//} \approx C_{\perp} = N\,g_{\text{av}}^2\,\beta^2 S(S+1)/3k_{\text{B}}, \tag{9}$$

where $g_{\text{av}}$ is the "average" g-factor that reproduces the slopes of the $1/\chi_i^{\text{eff}}$ vs. $T$ plots. Then, from Eqs. 7 and 9 we obtain

$$\theta_i \approx zJ\,(g_{\text{av}}/g_i)^2 S(S+1)/3k_{\text{B}} \qquad (i = //, \perp), \tag{10}$$

which leads to

$$\frac{\theta_{//}}{\theta_{\perp}} \approx \left(\frac{g_{\perp}}{g_{//}}\right)^2. \tag{11}$$

This expression shows that the Weiss temperature should be larger in magnitude along the direction of the smaller g-factor.

To justify the above modification of the Curie-Weiss law, we note the expression of the anisotropic molar susceptibility Lines derived for the $Co^{2+}$ ions of $CoCl_2$,[16]

$$\chi_i = \frac{4N\beta^2(A_i + B_i k_{\text{B}} T / \lambda)}{4k_{\text{B}} T + 6J(1 + \sigma_i \alpha)}, \qquad (i = //, \perp) \tag{12}$$

where $\lambda$ is the effective spin-orbit coupling constant for $CoCl_2$, $\sigma_\perp = 0$ and $\sigma_{//} = 1$, and

$$\alpha = 1 - \left(\frac{g_{//}}{g_\perp}\right)^2.$$  (13)

The parameters $A_i$ and $B_i$ are discussed below. If we set Eq. 12 equal to the modified Curie-Weiss law, Eq. 7, we find

$$C_{//} = \frac{N\beta^2}{k_B}(A_{//} + B_{//}k_BT/\lambda),$$  (14a)

$$C_\perp = \frac{N\beta^2}{k_B}(A_\perp + B_\perp k_BT/\lambda),$$  (14b)

$$\theta_{//} = \frac{6J}{4k_B}(1+\alpha),$$  (15a)

$$\theta_\perp = \frac{6J}{4k_B}.$$  (15b)

The parameter $A_i$ is proportional to the thermally-averaged $g_i^2$, while $B_i$ is the thermally-averaged second-order Zeeman coefficient.[16] Plots of the parameters $A_i$ and $B_i$ as a function of $\lambda/k_BT$ (**Figure 3** of Ref. 16) indicate that, in the high temperature region, $B_{//}$ and $B_\perp$ are proportional to $\lambda/k_BT$. The terms $B_ik_BT/\lambda$ ($i = //$, $\perp$) in the numerator of Eq. 12 are the slopes of the $B_i$ vs. $\lambda/k_BT$ curves, which reveals that $B_\perp k_BT/\lambda$ is greater than $B_{//}k_BT/\lambda$ (i.e., 5.47 vs. 4.53). In high temperature region, however, $A_\perp$ is smaller than $A_{//}$ (1.28 vs. 2.21) such that $A_{//} + B_{//}k_BT/\lambda = A_\perp + B_\perp k_BT/\lambda$ ($\approx 6.75$) and hence $C_{//} = C_\perp$. Furthermore, we obtain from Eqs. 13 and 15

$$\frac{\theta_{//}}{\theta_\perp} = 2 - \left(\frac{g_{//}}{g_\perp}\right)^2.$$  (16)

This expression is different from Eq. 11, but similar results are obtained from both expressions (see below).

## 4. Anisotropic magnetic susceptibilities of $Na_xCoO_2$

Single-crystalline $Na_{0.78}CoO_2$ was grown in an optical floating-zone furnace. Sizeable single crystals of approximate dimensions $2 \times 2 \times 0.5$ mm$^3$ were obtained by cleaving from the ingot. Their chemical composition was analyzed by inductively coupled plasma atomic emission spectroscopy. The c-axis lattice parameter of 10.765 Å, obtained from single crystal x-ray diffraction measurements along (00l) at room temperature, agrees well with the literature value for samples with comparable Na content.[19] The homogeneity of the magnetic properties was further confirmed by muon spin rotation (μSR) measurements, which provide direct evidence for the occurrence of a bulk long-range magnetic order below 22 K.[20] The purely nonmagnetic signal (except for nuclear contributions) above 22 K also indicates the absence of any significant amount of magnetic impurity phases. The electronic properties of the crystal were further characterized by infrared ellipsometry measurements, which exhibit a strongly temperature-dependent weakly-metallic behavior as well as a spin-dependent polaronic behavior in agreement with the report for $Na_{0.82}CoO_2$.[21] The anisotropic magnetic susceptibilities $\chi_{//}$ and $\chi_{\perp}$ were measured using a MPMS SQUID magnetometer (Quantum Design) in a field of 1 $T$.

The $\chi_{//}$ vs. T and $\chi_{\perp}$ vs. T plots of $Na_{0.78}CoO_2$ are presented in **Figure 1**. These plots were analyzed by using Eq. 8, without including a temperature independent contribution as a fitting parameter. As shown in the inset of **Figure 1**, the curves of $1/\chi_{//}$ vs. T and $1/\chi_{\perp}$ vs. T are quite linear in the $50 - 300$ K region. The linear parts of these plots lead to the values of $C_{//}$, $\theta_{//}$, $C_{\perp}$ and $\theta_{\perp}$ listed in **Table 1**, which reveals that $C_{//} \approx C_{\perp}$.

The value of $C_{//} \approx C_{\perp} \approx 0.21$ for $Na_{0.78}CoO_2$ was calculated under the assumption of 0.22 moles of $Co^{4+}$ ions per formula unit. Thus, for one mole of $Co^{4+}$ ions, $C_{//} \approx C_{\perp} \approx 0.95$ results. Using these values of the Curie constants and $S = 1/2$ for a LS $Co^{4+}$ ion in Eq. 9, we find $g_{av} \approx 3.18$. Note that $\theta_{//}/\theta_{\perp} \approx 245/144 \approx 1.7$, while $(g_{\perp}/g_{//})^2 \approx (4.84/3.38)^2 \approx 2.0$. Thus, the relationship $\theta_{//}/\theta_{\perp} \approx (g_{\perp}/g_{//})^2$ of Eq. 11 is supported. Moreover, $2 - (g_{//}/g_{\perp})^2 \approx 1.5$, so that the relationship $\theta_{//}/\theta_{\perp} \approx 2 - (g_{//}/g_{\perp})^2$ of Eq. 16 also holds.

For comparison the $\chi_{//}$ vs. T and $\chi_{\perp}$ vs. T data of $Na_{0.75}CoO_2$ reported by Chou *et al.*[9] and Sales *et al.*[10] were also analyzed by using Eq. 8, the results of which are summarized in **Table 1**. As expected, the values of $C_{//}$ and $C_{\perp}$ are approximately equal. The Weiss temperature is less negative for the $\perp$c-direction than for the //c-direction (i.e., $\theta_{//} < \theta_{\perp} < 0$) for both $Na_{0.78}CoO_2$ and $Na_{0.75}CoO_2$, and their $\theta_{\perp}/\theta_{//}$ ratios lie in between $(g_{//}/g_{\perp})^2$ and $2 - (g_{//}/g_{\perp})^2$. The substantially negative Weiss temperatures $\theta_{//}$ and $\theta_{\perp}$ of $Na_xCoO_2$ ($x \approx 0.75$) suggest the presence of strong predominant AFM spin exchange interactions. Since the interlayer Co-O…O-Co super-superexchange should be negligible to a first approximation,[14] this means that the intralayer spin exchange should be strongly AFM. As already mentioned, however, the neutron scattering studies [11,12] show that the intralayer spin exchange is FM while the interlayer spin exchange is strongly AFM. How to resolve these conflicting pictures will be discussed in Section 5. In the remainder of this section, we will assume that the occurrence of strongly AFM spin exchange in $Na_xCoO_2$ ($x \approx 0.75$), deduced from the magnetic susceptibility measurements in the temperature region of ~50 – 300 K, arises from the intralayer spin exchange.

The concentration of $Co^{4+}$ ions in $Na_xCoO_2$ (x ≈ 0.75) is low, and the susceptibility curves show a broad maximum at ~50 K (see **Figure 1**), which is characteristic of low-dimensional short range magnetic ordering. Therefore, if only the $Co^{4+}$ ions are responsible for the observed AFM behaviors of $Na_xCoO_2$, it would be necessary for the $Co^{4+}$ ions to form low-dimensional aggregates, e.g., chain fragments or rings with $z = 2$. Then, $J$ refers to the NN spin exchange interaction in such low-dimensional aggregates. On the basis of Eq. 10, the value of $J$ for $Na_{0.78}CoO_2$ can be estimated by using $z = 2$, $\theta_{//} = -245$ K, $\theta_{\perp} = -144$ K, $g_{//} = 3.38$, $g_{\perp} = 4.84$, $S = 1/2$ and $g_{av}$ ≈ 3.18. Thus, $J/k_B ≈ -550$ K for the $//c$-direction, and $J/k_B ≈ -670$ K for the $//\perp$-direction. The value of $J$ for $Na_{0.78}CoO_2$ can also be estimated by using Eq. 15, which are relevant for $z = 6$. These expressions lead to $J/k_B ≈ -330$ K for the $//c$-direction, and $J/k_B ≈ -290$ K for the $//\perp$-direction. Thus, the value of $J/k_B$ estimated from Eq. 10 is greater in magnitude than that found from Eq. 15 by a factor of about two. This suggests that the $z$ value should be greater than 2. In any case, what is clear is that the intralayer spin exchange is quite strong in $Na_{0.78}CoO_2$. Though not shown, the same is found for $Na_{0.75}CoO_2$ using the values of $\theta_{//}$ and $\theta_{\perp}$ listed in **Table 1**.

## 5. Discussion

So far, we have not considered the possibility that the $CoO_6$ octahedra of the $Co^{3+}$ ($d^6$) ions of $Na_xCoO_2$ surrounding the $Co^{4+}$ ions might undergo a slight change in geometry, and that their preferred spin state may deviate from the diamagnetic state resulting from the electron configuration $(t_{2g})^6$ expected for the undistorted environment. Consider the six $Co^{3+}$ ions in the immediate vicinity of a given $Co^{4+}$ ion in $Na_xCoO_2$ (**Figure 2**). In a $CoO_6$ octahedron, a $Co^{4+}$ ion requires shorter Co-O bonds than does a $Co^{3+}$ ion. A given $Co^{4+}$ ion can satisfy this requirement by shortening its six Co-O bonds. This will, in turn, lengthen those Co-O bonds of every $Co^{3+}$ ion that are connected to the oxygen atoms surrounding each $Co^{4+}$ ion. For simplicity of our discussion, the six $Co^{3+}$ ions surrounding a $Co^{4+}$ ion will be referred to as the NN $Co^{3+}$ ions. This Co-O bond lengthening will reduce the extent of the Co-O antibonding in, and hence lowers the energy of, the $e_g$ levels of the NN $Co^{3+}$ ions. Then, it becomes possible that the NN $Co^{3+}$ ions may adopt the IS electron configuration $(t_{2g})^5(e_g)^1$ with $S = 1$, as has been proposed in the literature.[21,22] Each $CoO_6$ octahedron containing a NN $Co^{3+}$ ion will lose the 3-fold rotational axis due to the unsymmetrical distortion associated with the Co-O bond lengthening (**Figure 2**). Therefore, its $t_{2g}$ and $e_g$ levels will each split. For convenience, the orbital designations $t_{2g}$ and $e_g$ will be used for the distorted $CoO_6$ octahedra containing NN $Co^{3+}$ ions because the extent of the distortion would be small. The $t_{2g}$ level occupation of an IS $Co^{3+}$ ion, i.e., $(t_{2g})^5$, is the same as that of a LS $Co^{4+}$ ion or a HS $Co^{2+}$ ion. Therefore, the NN $Co^{3+}$ ions are expected to provide anisotropic magnetic properties.

The above discussion implies that the actual mole fraction of the anisotropic spin sites in $Na_xCoO_2$ can be considerably larger than $(1−x)$ due to the presence of NN $Co^{3+}$ ions. If the $Co^{4+}$ ions are similar to their NN $Co^{3+}$ ions in their magnetic properties, they will form patches of geometrically frustrated spin arrangement [23] in which a given spin site may have as many as six neighbor sites (i.e., up to $z = 6$). This reduces the magnitude of the spin exchange $J/k_B$ value by a factor up to three (i.e., down to approximately $−200$ K from Eq. 10, which is more consistent with about $−300$ K from Eq. 15).

The existence of a large number of active magnetic spin sites in $Na_xCoO_2$, much higher than expected from the charge neutrality requirement, as well as the presence of different spin sites (e.g., the LS $Co^{4+}$ and IS $Co^{3+}$ sites) are consistent with the µSR experiments on $Na_{0.82}CoO_2$ [7] and $Na_{0.78}CoO_2$.[20] It also explains why $Na_xCoO_2$ (x ≈ 0.75) can undergo a three-dimensional (3D) AFM ordering below ~22 K because the $Co^{4+}$ ions and their NN $Co^{3+}$ ions may become connected and cover the whole trigonal lattice of each $CoO_2$ layer. As an illustration, an ordered arrangement of the $Co^{4+}$ ions in an isolated $CoO_2$ layer of $Na_{0.75}CoO_2$ is depicted in **Figure 2**. The possible presence of a large number of the IS $Co^{3+}$ ions in $Na_xCoO_2$ (0.5 < x < 1) also allows one to understand why these nonstoichiometric compounds are highly magnetic as well as weakly metallic.[4,7] As already discussed, the anisotropic magnetic properties of $Na_xCoO_2$ are explained by considering the $(t_{2g})^5$ electron configuration of the $Co^{4+}$ and their NN $Co^{3+}$ sites. Each IS $Co^{3+}$ site has one electron in the $e_g$ level. The $e_g$ levels of all $CoO_6$ octahedra, regardless of whether they contain $Co^{4+}$ or $Co^{3+}$ ions, overlap well to form wide $e_g$-block bands, which become partially filled due to the IS $Co^{3+}$ ions. This then explains the observed metallic properties of nonstoichiometric $Na_xCoO_2$ (0.5 < x < 1).

It is of interest to note the dependence of the Curie constants and the Weiss temperatures on the probing magnetic field. Sales *et al.*[10] reported magnetic susceptibility measurements of $Na_{0.75}CoO_2$ at $H$ = 0.1 and 5 $T$. Our analysis of their susceptibility data shows that the Curie constants and the Weiss temperatures decrease slightly in magnitude as the probing magnetic field increases (**Table 1**). This is a puzzling observation, but one might speculate whether it is caused by a very small amount of ferromagnetic impurities in single crystal samples. To prove this point, it would be necessary to carry out systematic magnetic susceptibility measurements as a function of the probing magnetic field. In any event, it should be emphasized that the main conclusions of our work are not affected by the above observation.

Finally, we discuss why the magnetic susceptibility and neutron scattering measurements of $Na_xCoO_2$ (x ≈ 0.75) present conflicting magnetic pictures. It is important to recall that the Weiss temperatures of $Na_xCoO_2$ are determined from the linear part of the $1/\chi_i$ vs. T plots ($i = \perp, //$) well above the 3D magnetic ordering temperature of ~22 K (i.e., between 50 − 300 K), whereas the neutron scattering experiments are performed well below that temperature (i.e., below ~2 K). For x ≈ 0.75, almost the whole trigonal lattice of each $CoO_2$ layer is covered by the $Co^{4+}$ and NN $Co^{3+}$ ions. The most probable arrangement is close to the ordered arrangement shown for x = 0.75 in **Figure 2**, namely, the NN $Co^{3+}$ ions form a honeycomb lattice in which the $Co^{4+}$ ions occupy the hexagon centers. (The arrangement of the $Co^{4+}$ and $Co^{3+}$ ions should be intimately related to that of the $Na^+$ ions. To minimize electrostatic repulsion, $Na^+$ ions should be located closer to the $Co^{3+}$ than the $Co^{4+}$ ions.) In the remainder of this work, we will assume that patches of such an arrangement of $Co^{4+}$ and NN $Co^{3+}$ ions cover each trigonal lattice of $CoO_2$ in $Na_xCoO_2$ when x ≈ 0.75. To explain the metallic property of $Na_xCoO_2$, the $e_g$ electron of each IS $Co^{3+}$ ion should be slightly delocalized, which implies that the average spin $\langle S \rangle$ of each IS $Co^{3+}$ ion would be lower than, but considerably greater than that of each LS $Co^{4+}$ ion. If we assume that the NN spin exchange parameter $J$ between adjacent LS $Co^{4+}$ and IS $Co^{3+}$ ions is the same as that between adjacent IS $Co^{3+}$ ions, the NN spin exchange energy $-J\hat{S}_i \cdot \hat{S}_j$ between two spin sites $i$ and $j$ should favor an AFM coupling between adjacent IS $Co^{3+}$ ions than between LS $Co^{4+}$ and IS $Co^{3+}$ ions. This will weaken the geometric spin frustration expected when magnetic ions cover a trigonal lattice, and lead the honeycomb lattice of IS $Co^{4+}$ ions to have an AFM coupling between all adjacent ions. As a consequence, the spin exchange interaction of each LS $Co^{4+}$ ion with its six NN neighbors will be effectively zero (i.e., three AFM plus three FM interactions) (**Figure 3**). The elastic neutron scattering measurements [11] of $Na_xCoO_2$ show the Bragg reflections below ~22 K that arise from ferromagnetically ordered $CoO_2$ planes with the average net moment of 0.13 $\mu_B$ per Co. The latter observation can be explained in terms of the FM ordering of the LS $Co^{4+}$ ions, because the IS $Co^{3+}$ ions of the honeycomb lattice will not contribute to these peaks because of their AFM coupling. Then the average effective moment expected from the elastic neutron scattering experiment is 0.25 $\mu_B$ per Co for $Na_{0.75}CoO_2$,[11] which is reasonable given the experimentally observed value of 0.13 $\mu_B$ per Co. It is also consistent with the heat capacity study of $Na_xCoO_2$,[7] which shows that a small fraction (i.e., ~10 %) of the total available spins are involved in the long range AFM ordering below ~22 K and the remainder is removed in short range ordering above ~22 K.

The neutron scattering experiments [11,12] show that the intralayer spin exchange ($J_\perp$) and the interlayer spin exchange ($J_{//}$) are quite substantial, and the interlayer spin exchange is stronger, in magnitude (i.e., $J_\perp = 3.3$ meV and $J_{//} = -4.5$ meV in Ref. 11; $J_\perp = 6.0$ meV and $J_{//} = -12.2$ meV in Ref. 12). Since the $Co^{4+}$ ions are well separated from each other within and between $CoO_2$ layers, these strong interactions are difficult to understand in terms of the usual spin exchange interactions resulting from localized spins, i.e., superexchange interactions involving Co-O-Co paths and super-super exchange interactions involving Co-O…O-Co paths.[14] The spin wave dispersion of $Na_xCoO_2$ (x ≈ 0.75) observed from the neutron scattering studies [11,12] are well reproduced by first-principles electronic band structure calculations.[24] It is most likely that the strong spin exchange interactions of $Na_xCoO_2$ are mediated by the delocalized electrons of the $e_g$ block bands. The intralayer FM coupling can be explained if the $e_g$ block bands of each $CoO_2$ layer have a half-metallic character, because a FM coupling between the spins of the LS $Co^{4+}$ ions with those of the conduction electrons leads effectively to an FM coupling between the spins of the LS $Co^{4+}$ ions. This is quite likely because the electronic band structure calculations [24] for the FM state show a half-metallic character. (In other words, the intralayer FM coupling is mediated by a RKKY mechanism.[25] One might consider double-exchange [26] interactions between adjacent LS $Co^{4+}$ and IS $Co^{3+}$ ions as a cause for the intralayer FM coupling. However, this double-exchange mechanism cannot explain the strong predominant AFM spin exchange seen by the magnetic susceptibility measurements nor the very small effective magnetic moment on Co seen by the neutron scattering experiments.) In addition, an interlayer AFM coupling would be energetically more favorable than an interlayer FM coupling, because the wave function of the AFM state has nodes in between the $CoO_2$ layers thereby preventing the $CoO_2$ layers from transferring some valence electron density from the $CoO_2$ layers into the region between them.[27]

## 6. Concluding remarks

For a magnetic system made up of anisotropic magnetic ions the Curie-Weiss law is modified as in Section 3. The anisotropic magnetic susceptibilities of $Na_xCoO_2$ (x ≈ 0.75) are well described by the modified Curie-Weiss law without including the temperature independent contribution. Our analysis shows that $C_{//} \approx C_{\perp}$, and the Weiss temperature is more negative along the direction of the lower g-factor (i.e., $\theta_{//} < \theta_{\perp} < 0$, and $g_{//} < g_{\perp}$). The LS $Co^{4+}$ ($S = 1/2$) ions of $Na_xCoO_2$ generated by the Na vacancies lead to anisotropic magnetic properties as do the HS $Co^{2+}$ ($S = 3/2$) ions of $CoCl_2$. The metallic and anisotropic magnetic properties of $Na_xCoO_2$ are accounted for if it is assumed that the six $Co^{3+}$ ions surrounding each $Co^{4+}$ ion adopt the IS electron configuration $(t_{2g})^5(e_g)^1$. Our analysis suggests that the magnetic properties of $Na_xCoO_2$ ($x \approx 0.75$) deduced from the magnetic susceptibilities in the temperature region of ~50 – 300 K should arise from the intralayer spin exchange interactions associated with the IS $Co^{3+}$ and LS $Co^{4+}$ ions. The magnetic susceptibility and neutron scattering studies of $Na_xCoO_2$ present apparently contradicting pictures of the magnetic structure. This conflict is resolved by noting that the AFM spin exchange between adjacent IS $Co^{3+}$ ions is stronger than that between adjacent IS $Co^{3+}$ and LS $Co^{4+}$ ions, so an AFM coupling takes place between all adjacent IS $Co^{3+}$ ions of the honeycomb lattice when the temperature is lowered toward ~22 K. The spin wave dispersion seen from the inelastic neutron scattering studies below ~2 K as well as the Bragg reflections of ferromagnetically ordered $CoO_2$ planes below ~22 K should be related to the intralayer FM and interlayer AFM coupling between the LS $Co^{4+}$ ions mediated by the delocalized conduction electrons of the partially filled $e_g$ bands.

**Acknowledgments**

The research at North Carolina State University was supported by the Office of Basic Energy Sciences, Division of Materials Sciences, U. S. Department of Energy, under Grant DE-FG02-86ER45259. We thank Professor C. Bernhard for drawing our attention to the anisotropic magnetic susceptibilities of $Na_xCoO_2$, and Dr. S. P. Bayrakci and Professor C. Bernhard for providing us with experimental data prior to publication and also for useful discussions.

**References**

(1)     Fouassier, C.; Matejka, G.; Reau, J.-M.; Hagenmuller, P., *J. Solid State Chem*. **1973**, *6*, 532.

(2)     Jansen, M.; Hoppe, R., *Z. Anorg. Allg. Chem*. **1974**, *408*, 104.

(3)     Takahashi, Y.; Gotoh, Y.; Akimoto, J., *J. Solid State Chem*. **2003**, *172*, 22.

(4)     Delmas, C.; Braconnier, J. J.; Fouassier, C.; Hagenmuller, P., *Solid State Ionics* **1981**, *3/4*, 165.

(5)     Takada, K.; Sakurai, H.; Takayama-Muromachi, E.; Izumi, F.; Dilanian, R.; Sasaki, T., *Nature (London)* **2003**, *422*, 53.

(6)     Gavilano, J. L.; Rau, D.; Pedrini, B.; Hinderer, J.; Ott, H. R.; Kazakov, S. M.; Karpinski, J., *Phys. Rev. B* **2004**, *69*, 100404.

(7)     Bayrakci, S. P.; Bernhard, C.; Chen, D. P.; Keimer, B.; Kremer, R. K.; Lemmens, P.; Lin, C. T.; Niedermayer, C.; Strempfer, J., *Phys. Rev. B* **2004**, *69*, 100410(R).

(8)     Zandbergen, H. W. Foo, M.; Xu, Q.; Kumar, V.; Cava, R. J., *Phys. Rev. B* **2004**, *70*, 024101.

(9)     Chou, F. C.; Cho, J. H.; Lee, Y. S., *Phys. Rev. B* **2004,** *70*, 144526.

(10) Sales, B. C.; Jin, R.; Affholter, K. A.; Khalifah, P.; Veith, G. M.; Mandrus, D., *Phys. Rev. B* **2004**, *70*, 174419.

(11) Bayrakci, S. P.; Mirebeau, I.; Bourges, P.; Sidis, Y.; Enderle, M.; Mesot, J.; Chen, D. P.; Lin, C. T.; Keimer, B., *Phys. Rev. Lett*. **2005**, 64, 157205.

(12) Helme, L. M.: Boothroyd, A. T.; Coldea, R.; Prabhakaran, D.; Tennant, D. A.; Hiess, A.; Kulda, J., *Phys. Rev. Lett*. **2005**, 64, 157206.

(13) Boothroyd, A. T.; Coldea, R.; Tennant, D. A.; Prabhakaran, D.; Helme, L. M.; Frost, C. D., *Phys. Rev. Lett*. **2004**, *92*, 197201.

(14) For recent reviews see: (a) Whangbo, M.-H.; Koo, H.-J.; Dai, D., *J. Solid State Chem*. **2003**, *176*, 417. (b) Whangbo, M.-H.; Dai, D.; Koo, H.-J., *Solid State Sci*. **2005**, *7*, 827.

(15) (a) Gašparović, G.; Ott, R. A.; Cho, J.-H.; Chou, F. C.; Chu, Y.; Lynn, J. W.; Lee, Y. S., *Phys. Rev. Lett*. **2006**, *96*, 046403.

(b) Choy, T.-P.; Galanakis, D.; Phillips, P., *arXiv:cond-mat/0502164*.

(16) Lines, M. E. *Phys. Rev*. **1963**, *131*, 546.

(17) Dai, D.; Whangbo, M.-H., *Inorg. Chem*. **2005**, *44*, 4407.

(18) Smart, J. S., *Effective Field Theories of Magnetism*, W. B. Saunders Company, Philadelphia, 1966.

(19) Huang, Q.; Foo, M. L.; Pascal, Jr., R. A.; Lynn, J. W.; Toby, B. H.; He, T.; Zandbergen, H. W.; Cava, R. J., *Phys. Rev. B* **2004**, *70*, 184110.

(20) Bernhard, C. *et al*., in preparation.

(21) Bernhard, C.; Boris, A. V.; Kovaleva, N. N.; Khaliullin, G.; Pimenov, A. V.; Yu, L.; Chen, D. P.; Lin, C. T.; Keimer, B., *Phys. Rev. Lett*. **2004**, *93*, 167003.

(22) Khaliullin, G., *arXiv: cond-mat/0510025*.

(23) (a) Greedan, J. E., *J. Mater. Chem*. **2001**, *11*, 37. (b) Dai, D.; Whangbo, M.-H., *J. Chem. Phys*. **2004**, *121*, 672.

(24) Johannes, M. D.; Mazin, I. I.; Singh, D. J., *Phys. Rev. B* **2005**, *71*, 214410.

(25) (a) Rudemann, M. A.; Kittel, C., *Phys. Rev*. **1954**, *96*, 99. (b) Kasuya, T., *Prog. Theoret. Phys*. **1956**, *16*, 45. (c) Yoshida, K., *Phys. Rev*. **1957**, *106*, 893.

(26) Zener, C., *Phys. Rev*. **1951**, *82*, 403.

(27) Villesuzanne, A.; Whangbo, M.-H., *Inorg. Chem*. **2005**, *44*, 6339.

Table 1.    Fitting parameters of the anisotropic magnetic susceptibilities $\chi_{//}$ and $\chi_{\perp}$ of $Na_xCoO_2$ (x = 0.78 and 0.75) by the modified Curie-Weiss law [a]

| x | $H$ | $C_{//}$ | $\theta_{//}$ | $C_{\perp}$ | $\theta_{\perp}$ | $\theta_{//}/\theta_{\perp}$ |
|---|---|---|---|---|---|---|
| 0.78 [b] | 1 | 0.208 | -245 | 0.214 | -144 | 1.70 |
| 0.75 [c] | 0.1 | 0.380 | -420 | 0.387 | -279 | 1.51 |
| 0.75 [d] | 5 | 0.315 | -332 | 0.347 | -206 | 1.61 |
| 0.75 [e] | 1 | 0.322 | -354 | 0.310 | -223 | 1.59 |

[a] $H$ is in units of $T$, $\theta_{//}$ and $\theta_{\perp}$ are in units of K, and $C_{//}$ and $C_{\perp}$ are in units of $cm^3$ K/mol.

[b] This work        [c] Ref. 10        [d] Ref. 10        [e] Ref. 9

**Figure caption**

Figure 1.      $\chi_{//}$ vs. T and $\chi_{\perp}$ vs. T plots measured for $Na_{0.78}CoO_2$. The inset shows the

$1/\chi_{//}$ vs. T and $1/\chi_{\perp}$ vs. T plots.

Figure 2.      Ordered arrangement of the $Co^{4+}$ ions in an isolated $CoO_2$ layer of

$Na_{0.75}CoO_2$. The large blue circles represent the $Co^{4+}$ ions, the large yellow circles

the NN $Co^{3+}$ ions, and the small red circles the O atoms attached to the $Co^{4+}$ ions.

To emphasize the site symmetry of the NN $Co^{3+}$ ions, the O atoms linked only to the NN $Co^{3+}$ ions are represented by small yellow circles.

Figure 3.      Model for the ferromagnetic $CoO_2$ planes of $Na_{0.75}CoO_2$ below ~22 K.

The honeycomb lattice of IS $Co^{3+}$ spins has an AFM coupling between adjacent ions, while the LS $Co^{4+}$ spins located at the hexagon centers are ferromagnetically ordered. The blue and red circles represent up-spin and down-spin sites, respectively.

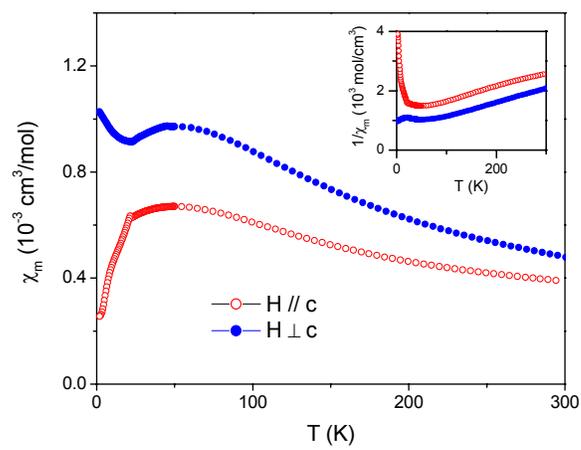

Figure 1

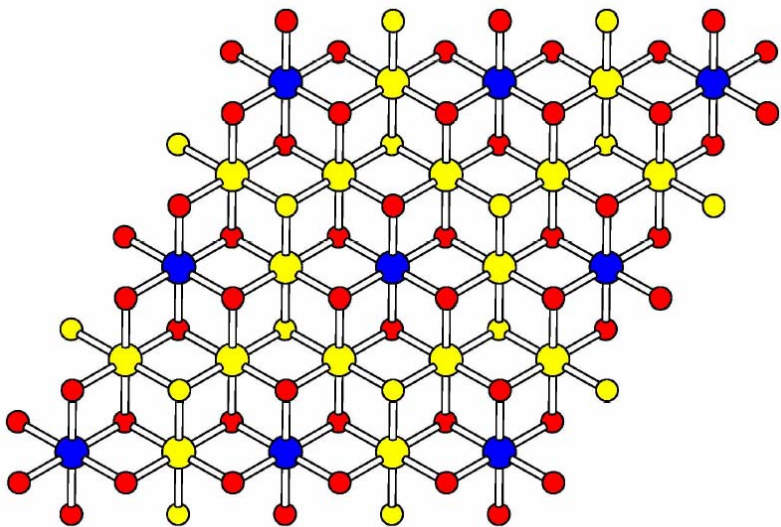

Figure 2

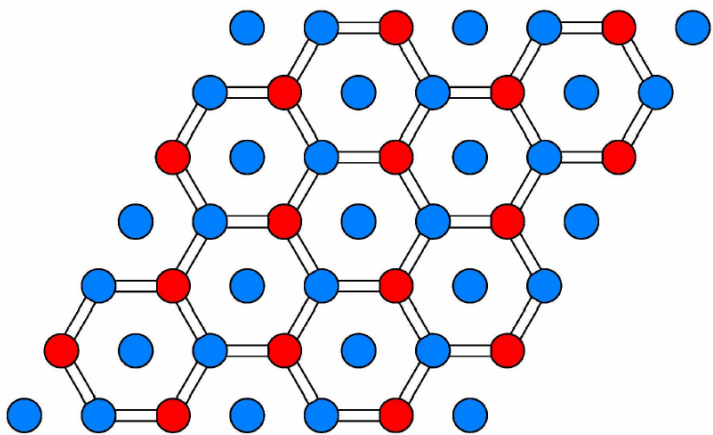

Figure 3

**Synopsis**

A modified Curie-Weiss law was proposed to analyze the anisotropic magnetic susceptibilities of $Na_xCoO_2$ ($x \approx 0.75$), and implications of this analysis concerning the metallic properties and the long-range antiferromagnetic ordering of $Na_xCoO_2$ were explored. The magnetic structures of $Na_xCoO_2$ deduced from neutron scattering and magnetic susceptibility studies were shown to be in apparent contradiction, and how to resolve this conflict was probed.

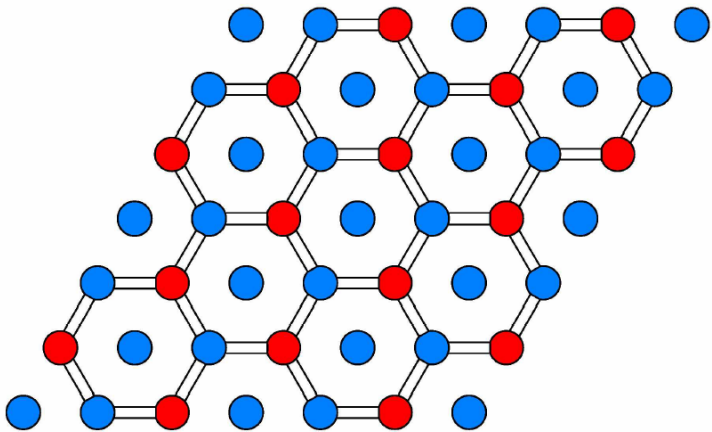